\documentclass[peerreview,12pt]{IEEEtran}
\IEEEoverridecommandlockouts
\usepackage{cite}
\usepackage{amsmath,amssymb,amsfonts}
\usepackage{algorithmic}
\usepackage{graphicx}
\usepackage{textcomp}
\usepackage{xcolor}
\usepackage{setspace} 
\def\BibTeX{{\rm B\kern-.05em{\sc i\kern-.025em b}\kern-.08em
    T\kern-.1667em\lower.7ex\hbox{E}\kern-.125emX}}
\begin{document}

\title{UAV-Enabled IoT Networks: A SWIPT Energy Harvesting Architecture with Relay Support for Disaster Response\\
{\footnotesize \textsuperscript{}}
}

\author{\IEEEauthorblockN{Hossein Mohammadi Firouzjaei,\ \ Javad Zeraatkar Moghaddam,\ \ Mehrdad Ardebilipour}}

\maketitle

\begin{abstract}
Due to the wide application of unmanned aerial vehicles (UAVs) as relays to establish Disaster Response Networks (DRNs), an effective model of energy harvesting (EH) and energy consumption for the UAV-aided Disaster Response Network (DRN) is rising to be a challenging issue. This is mainly manifest in Internet of Things (IoT) scenarios where multiple users are looking to communicate with the UAV. In this paper, the possibility of connecting an UAV with several users is investigated where the UAV as a relay receives data from a DRN and delivers to another network considering two IoT scenarios. The first scenario represents a conventional method with limited UAV energy where low communication rates and inadequate service coverage for all users are challenges. But in the second scenario, a Simultaneous Wireless Information and Power Transmission (SWIPT) technique is used to serve users. Considering potential limitations in transmission energy of users within disaster networks, the SWIPT technique is applied to maximize energy acquisition by the UAV, leading to improve the efficiency of the investigated scenario. Finally, the required energy of the UAV to serve the largest number of users in the shortest possible time is clarified. Furthermore, by Considering the relationship between energy and UAV flight time and defining the UAV flight time optimization problem, optimal network parameters are obtained. Simulation results show the effectiveness of the proposed scenario.
\end{abstract}
\begin{IEEEkeywords}
UAVs, \ DRNs, \ SWIPT, \ EH
\end{IEEEkeywords}

\section{Introduction}
Drones, also known as UAVs, have gained significant attention as Flying Base Stations (FBSs) due to their ease of implementation, cost-effectiveness, and direct communication capabilities with users. However, their potential role in 5G telecommunications and beyond extends far beyond FBS applications. Drones offer the ability to enhance communication quality, expand coverage areas, and increase communication network capacity [1], [2]. Notably, the unique Line-of-Sight (LOS) communication capability of UAVs, combined with their mobility and flexibility, opens up opportunities for their utilization in establishing temporary communication networks, particularly Drone Relay Networks (DRNs).

The integration of Simultaneous Wireless Information and Power Transfer (SWIPT) technologies in DRNs presents a promising approach to improve network efficiency. Since UAVs possess limited energy and must collect data within a restricted timeframe [2], [3], SWIPT offers a solution to enhance UAV energy efficiency. The energy level of a UAV directly impacts its flight time, making it crucial to minimize the UAV’s mission duration to conserve energy. Recent studies have shown promising results in addressing the challenges of limited UAV battery energy and flight time minimization. For example, Viet-Hung Dang et al. [4] proposed a communication protocol incorporating an EH phase and multiple communication phases. UAV relays gather energy during the EH phase from a power source, subsequently serving as relays for secure information transmission dur-ing other telecommunication phases. Through optimization of relay drone positioning and determination of primary and secondary relays, resource allocation challenges are resolved, ensuring efficient signal delivery to IoT ground users across the network.

Existing research on relay-assisted UAV-aided networks [5-11] has primarily focused on separate modeling of EH and data collection. While a few models consider network energy efficiency alongside communication rate improvement, none have specifically addressed UAV relay networks with EH and data reception in the uplink phase, followed by data transmission in the downlink phase, while minimizing UAV flight duration. Diverging from prior studies, this paper investigates the energy consumption issue within a scenario where the UAV traverses between two areas, simul-taneously engaging in data transmission and battery charging. By employing EH techniques and optimizing UAV parameters, the paper aims to minimize flight time and mitigate energy consumption challenges.

This paper explores a system, as illustrated in Figure 1, across two distinct scenarios, demonstrating the effectiveness of SWIPT in UAV-aided DRNs. The first scenario represents the conventional method of using drones in telecommunication networks, where the drone receives data from one network, moves to the central network or another node, and delivers the data. Due to limited UAV energy, this approach faces challenges such as low communication rates and inadequate service coverage for all users. In the second scenario, the UAV utilizes EH techniques to charge its battery while communicating with the first area, maximizing energy availability in the network. Considering potential limitations in transmission energy for users within disaster networks, the EH technique is applied to maximize energy acquisition by the UAV, leading to improved efficiency com-pared to the first scenario. Notably, this paper emphasizes that the UAV establishes both uplink and downlink communication with ground users (GUs), leveraging its unique LOS communication capability, which is extensively demonstrated.

Figure 1 depicts the process where the UAV receives data and subsequently moves from the right area to the left area, establishing communication with the base station (BS) in the latter area to fulfill its mission. Consequently, the UAV’s energy reserves become heavily compromised, potentially hin-dering its ability to fulfill its role as a relay between the two areas. To address this challenge, EH techniques are employed in the second scenario, enabling the UAV to complete its operations and serve a greater number of users.
\begin{figure}[!t]
\setlength{\tabcolsep}{4pt}
\centerline{\includegraphics[width=3in]{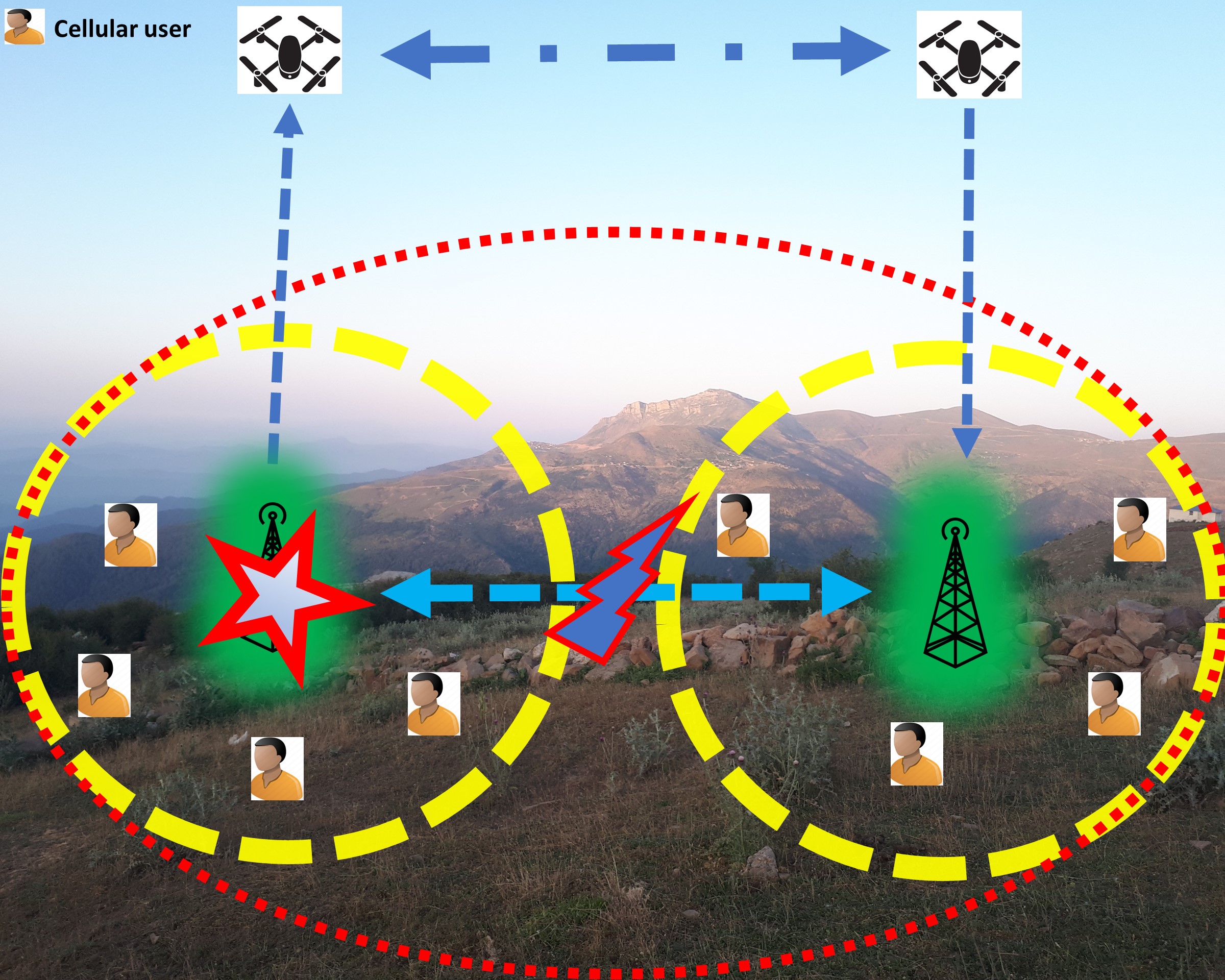}}
\caption{System Model}
\label{fig1}
\end{figure}

This paper estimates the energy consumed by the UAV and formulates an optimization problem focused on minimizing UAV flight time. As flight time directly influences energy consumption, minimizing it allows for maximum data acqui-sition with minimal energy usage. The optimization problem’s constraints ensure sufficient energy provision for the UAV to cover the area and serve all users, while maximizing users’ sending and receiving rates. By comparing two scenarios, A and B, this paper demonstrates the superior performance of the proposed EH-based scenario (B) leveraging EH techniques.

\subsection{Related Work}
The field of DRNs faces significant challenges due to human disasters such as fire or war, as well as natural catastrophes like floods or storms, which often render ground base stations unsuitable [12], [13]. In such scenarios, drones have emerged as crucial elements in DRN networks, providing reliable telecommunication con-nections to ground users. However, UAV-enabled communi-cation networks encounter a fundamental challenge, namely the limited flight time of drones, while ground users require stable and long-term communication. This limitation arises from the finite energy capacity of drone batteries, which act as their primary power source [14], [15]. As a result, recent studies have extensively explored techniques that can address the limited energy problem of UAVs and supply energy during communication. In the following section, we present relevant research conducted in this area.

Juan Zhang et al. [1] discussed the maintenance of energy efficiency in UAVs and the guarantee of reputation gain during scheduling deployment, particularly in the context of the Inter-net of Things (IoT). The authors formulated a UAV scheduling decision model that takes into account energy consumption and reputation. By utilizing game theory, they proposed an optimal decision search plan. The model enables the evaluation of drone scheduling parameters and facilitates the develop-ment of a suitable scheduling strategy that simultaneously increases drone reputation and reduces energy consumption. The application of game theory is crucial for solving the scheduling problem in drone communities when faced with numerous requests. Ultimately, the proposed model allows for a compromise between maximizing reputation and minimizing energy, efficiently achieving the Nash equilibrium for UAV scheduling and ensuring optimal scheduling for UAVs.

In [2], the authors presented a Markov Decision Process (MDP) model to optimize energy and data transmission for UAVs. The proposed model considers the exchange of energy and data between drones and Internet of Things users, with assistance from base stations to recharge the drone’s battery. To maximize the utilization of UAV flight time, the MDP model seeks a suitable strategy for data collection, delivery, and energy replenishment. The authors employed the value iteration algorithm to solve the model, and further applied Q-learning and deep reinforcement learning (DRL) schemes to address system state uncertainties and the large state space of UAV-assisted communication systems. The combined MDP model with DRL-based design achieves a more effective strategy for wireless energy and data transmission compared to traditional approaches. By solving the proposed MDP model, the UAV can complete its mission while considering network energy efficiency, data delivery delay, and the maximization of rewards received.

In [3], the authors investigated downlink communication in a multi-band heterogeneous network (HETNET) that utilizes two UAVs as flying base stations, along with several ground base stations serving nearby users. The study formulates a two-layer optimization problem, simultaneously determining a suitable antenna radius for UAVs and efficiently managing en-ergy resources. To achieve this, they enhance network energy efficiency (EE) by defining a SINR limit for users. The paper provides a comprehensive review of 5G networks, Line-of-Sight (LOS) communication, and the use of millimeter waves in 5G. The optimization problem incorporates constraints such as minimum Quality-of-Service (QOS) requirements and maximum communication power, ensuring acceptable commu-nication quality and maximizing data transfer.

The time spent by drones comprises two parts: communi-cation with users and movement between nodes [16], [17]. Similarly, the energy consumed by drones consists of two components: energy used for communication and energy ex-pended during movement [18]. While the energy spent on communication is negligible, the time spent on both tasks and the energy utilized for movement pose significant challenges [14], [15]. EH techniques have emerged as promising solutions to extend the service time of UAVs. By harnessing energy from users, the UAV’s energy level can be increased, thereby enhancing its flight time. Recent papers [19] have shown interesting results regarding EH techniques in downlink communications, and implementing this technique in uplink communications can be highly effective, particularly during UAV mobility.

Despite the advantages of increased coverage and the ability to position drones in the main lobe path of beamforming techniques, mobile drones significantly consume more energy [20], [21]. The EH technique becomes even more crucial when drones are in motion, as they consume a substantial amount of energy, even at high speeds. A promising solution to improve network communication quality lies in striking a trade-off between the advantages and disadvantages of UAV mobility, including increased flight duration and energy consumption [22-26].

R. S. de Moraes et al [27] have presented the development of an autonomous and distributed movement coordination algo-rithm for UAV swarms used in communication relay networks and exploratory area surveillance missions. The proposed hybrid algorithm combines pheromone maps, market auction paradigms, and proactive link maintenance mechanisms to create a self-organizing flying network. Simulations were conducted to validate the algorithm’s performance in target allocation and network connectivity. The results demonstrate that the proposed solution effectively balances performance goals. The algorithm addresses UAV movement planning, task distribution, and trade-off missions, considering network connectivity, area coverage, and handling tasks in Points of Interest (POIs). It utilizes a potential map algorithm inspired by ant-pheromone paradigm for movement coordination, a distributed beacon approach for network connectivity, and an auction-based approach for task allocation. The proposed solution offers coordinated movement, behavior, and mission handling for UAV groups, enabling exploration, connectivity management, and task allocation in unstable network environments. Simulations confirm the system’s capability to perform tasks effectively, albeit with a requirement for proper parameter calibration for optimal results.

To the best of our knowledge, no existing literature proposes a model that incorporates EH techniques in uplink communication and data transmission using the harvested energy, while accounting for UAV movement between areas. Furthermore, this paper presents a unique approach by comparing a scenario without EH techniques to one equipped with EH techniques, thereby demonstrating the benefits of implementing EH techniques in UAV-based service networks.

\subsection{Motivations and Key Contributions}
Motivations:
\begin{itemize}
  \item The wide application of UAVs as relays in DRNs necessitates an effective model of EH and energy consumption for UAV-aided DRNs.
  \item In Internet of Things (IoT) scenarios, where multiple users need to communicate with the UAV, the issue of providing sufficient energy to the relay UAV to serve the largest number of users in the shortest time is crucial.
  \item Existing research has modeled EH and data collection separately in relay-assisted UAV-aided networks, but there is a lack of studies that address the energy consumption issue of UAVs in scenarios where the UAV moves between areas, charges its battery, and minimizes flight duration.
\end{itemize}	

Key Contributions:
\begin{itemize}
  \item The paper compares two scenarios in UAV-aided DRNs, focusing on the effectiveness of SWIPT techniques. The first scenario represents the traditional method of using drones in telecommunication networks, while the second scenario utilizes EH techniques to charge the UAV’s battery while communicating with the first area, maximizing the available energy in the network.
  \item The paper formulates uplink and downlink communi-cation rates and energy consumption, and then applies SWIPT techniques to charge the UAV’s battery and optimize network parameters, specifically the UAV flight time.
  \item Theoretical results demonstrate that the proposed EH strategy achieves optimal performance for the UAV. Simulation results validate the theoretical findings.
  \item The paper highlights the importance of using SWIPT technologies to improve the efficiency of UAV-aided DRNs, leading to increased communication quality, ex-panded coverage, and enhanced capacity of communica-tion networks.
  \item The findings of the study show that the proposed SWIPT-based scenario outperforms the traditional scenario, en-abling the UAV to serve more users from the disaster area and minimizing energy consumption.
  \item Future work could explore the integration of multiple-input multiple-output (MIMO) systems with several UAVs assisted by beamforming and SWIPT technologies to form a network of IoT users in an area.
\end{itemize}

\subsection{Organization}
The rest of the paper is organized as follows. Section II illustrates the framework outline of system. After that, the EH mechanism for the UAV-aided DRN is studied in section III. Following that in section IV, the optimization problem is solved. Simulation results are shown in section V and also comparison between analytical results and simulation results are done in this section. Finally, section VI concludes the paper.

\section{Framework Outline}
One DRNs is established in the first area or the left area as shown in figure 1 and there is a base station in the second area or the left area as given in figure 1. The users of both networks are cellular users, and due to the failure of the Ground Base Station (GBS) of the first area, communication between the users of the two networks is impossible and a UAV is responsible to collect data from the first area and move to the second area to send them to the BS of the second area. Actually, the UAV plays the role of a flying base station (FBS) to serve the cellular users of the left network (NET1) by establishing an uplink connection and receives the data of the users of this network, and then by moving to the network on the right side (NET2) communicates with its BS in the form of downlink communications.  In fact, the UAV is a relay between NET1 and NET2, which connects these two networks to each other and conveys the information of the users of NET1 to the users of NET2.

During the time of communicating with NET1, the drone harvests energy from the users of NET1 in addition to their information, so that it has the necessary energy to fly to NET2 and communicate with the BS of NET2. The signal sent from NET1 users is received in the UAV and the energy of these signals is extracted and the battery of the drone is recharged. The UAV's uplink and downlink communication with both NET1 and NET2 is orthogonal, therefore, users' signals to the UAV does not interfere with each other.

Both scenarios A and B will be investigated completely and the main different between them is that the UAV harvests energy in scenario B. From now, each result or equation that includes EH techniques is related to scenario B and other equations are the same between two scenarios. In NET1, cellular users have a homogeneous Poisson Point Process (PPP) distribution $\rho_{c}^{N_{1}}$ with density $\lambda_{c}^{N_{1}}$($\ users/m^{2}$). There are two various channels in the proposed system, Ground to Air (G2A) channel and Air to Ground (A2G) channel. The G2A channel is established between the users of NET1 and the UAV. The A2G channel is established between the UAV and the BS of NET2.

Let use $P_{T}^{NET_{1}}$ as the transfer power of each uplink cellular user of NET1 that is fixed for all of them. Uplink signals that are received at the UAV include line-of-sight (LOS), strongly reflected non-LOS (NLOS), and multiple reflected components. Mentioned components are independent of each other and from environmental parameters like trees and buildings. The probability of LOS component for $k_{th}$ uplink cellular user of NET1 is indicated by $P_{LOS,k}^{NET_{1}}$ and the probability of NLOS component for $k_{th}$ uplink cellular user of NET1 is indicated by $P_{NLOS,k}^{NET_{1}}$; Where {$P_{LOS}$+$P_{NLOS}$}=1 for each user.

Therefore, the power of received signal from the $k_{th}$ uplink cellular user of NET1 at the UAV is computed as [6]:
\begin{equation}
    P_{R,k}^{UAV}=(P_{LOS,k}^{NET_{1}}+\eta P_{NLOS,k}^{NET_{1}})P_{T}^{NET_{1}}R_{k,UAV}^{-\alpha _{u}}
\end{equation}
Where, $\eta$ is the extra attenuation factor for the NLOS links. Then $P_{LOS,k}^{NET_{1}}$ is computed as follows:
\begin{equation}
    P_{LOS,k}^{NET_{1}}=\frac{1}{1+ Cexp(-B(\theta_{k}^{NET_{1}}-C))}\\
\end{equation}
\begin{equation}
    \theta_{k}^{NET_{1}}=\frac{180}{\pi}sin^{-1}(\frac{H^{UAV}}{R_{k,UAV}}\\)\\
\end{equation}

So, the Signal-Noise-Ratio(SNR) of the communication between the $k_{th}$ uplink cellular user of NET1 and the UAV is given as [6]:
\begin{equation}
    \gamma_{k}^{UAV}=\frac{P_{R,k}^{UAV}}{n_{NET1}^{UAV}}\\
\end{equation}
Where, $n_{NET1}^{UAV}$ is the power of additive white gaussian noise received from each uplink cellular user of the NET1 at the UAV.

Following that, the rate of the communication of $k_{th}$ uplink cellular user of NET1 with the UAV in the scenario A that has no EH technique, can be explained as:
\begin{equation}
    \zeta_{k}^{UAV}=Wlog_{2}(1+ \gamma_{k}^{UAV})
\end{equation}
Where $W$ is the bandwidth of the frequency band of $k_{th}$ uplink cellular user of NET1.

It will be assumed that UAV can receive a certain amount of data from each uplink user that is indicated by $D_{th}^{UAV}$, so the time that the UAV spends to collect data from the $k_{th}$ uplink cellular user of NET1 is computed as:
\begin{equation}
    T_{k}^{UAV}=\frac{D_{th}^{UAV}}{\zeta_{k}^{UAV}}\\
\end{equation}

There will be downlink communication between the UAV and the BS of NET2. The power of the received signal from the UAV in the BS of NET2 will be calculated almost like uplink users:
\begin{equation}
    P_{R,BS}^{UAV}=(P_{LOS,BS}^{NET_{2}}+\eta P_{NLOS,BS}^{NET_{2}})P_{T}^{UAV}H^{-\alpha _{u}}
\end{equation}
Where $P_{LOS,BS}^{NET_{2}}$ denotes the probability of LOS component for the BS of NET2 and $P_{NLOS,BS}^{NET_{2}}$ is the probability of NLOS component for the BS of NET2. $P_{T}^{UAV}$ is the transfer power of the UAV and $H$ indicates the altitude of the UAV on the above of NET2.

Similarly, the SNR, the rate of the communication between the UAV and the BS of NET2 and the time that the UAV has to spend to communicate during this period will be calculated as follows:
\begin{equation}
    \gamma_{BS}^{UAV}=\frac{P_{R,BS}^{UAV}}{n_{NET2}^{UAV}}\\
\end{equation}
\begin{equation}
    \zeta_{BS}^{UAV}=Wlog_{2}(1+ \gamma_{BS}^{UAV})
\end{equation}
\begin{equation}
    T_{BS}^{UAV}=\frac{D_{th}^{UAV}}{\zeta_{BS}^{UAV}}\\
\end{equation}
Where ${n_{NET2}^{UAV}}$ is the power of additive white gaussian noise received from the UAV in the BS of NET2.

The consumed energy by the UAV and the harvested energy of the UAV in scenario B will be calculated in the next section.

\section{EH Mechanism for the UAV-aided DRN}
In order to be able to transmit information and power simultaneously in telecommunication networks, SWIPT technology is a promising solution that has recently attracted the attention of many telecommunication network architects. However, due to different levels of sensitivity in both EH and information decoding (ID) operations, different receiver architectures are required to facilitate SWIPT. The received signal must be split into two separate parts, one for ID and one for EH. In general, EH and ID cannot be implemented on the same signal in the SWIPT system. The reason for this fact is that the implementation of EH on the radio frequency (RF) signal leads to the destruction of the information content of the signal [24], [25].

As a result, there are two ways to implement SWIPT:
\begin{itemize}
  \item Split the received signal into two parts
  \item Allocation of separate antennas for EH and ID
\end{itemize}

In general, 4 types of receiver architecture are used in SWIPT systems [24]:
\begin{itemize}
  \item Separate Receiver (SR): In this proposed design, EH and ID architectures are integrated, each equipped with a pair of independent receivers boasting separate antennas. These receivers operate in conjunction with a transmitter featuring multiple antennas. Importantly, the distinct antennas possess separate channels, providing enhanced flexibility. Remarkably, this receiver setup can be readily established utilizing off-the-shelf components, ensuring ease of implementation for both EH and ID receivers. To achieve simultaneous and independent componentization for ID and EH, we employ the SR architecture. It allows us to strike the right balance between extracted energy and information rate through receiver feedback and channel state information (CSI). The system's performance effectively can be optimized in this approach.
  \item Time Switching (TS):A common antenna for harvesting the energy and receiving data is shared in the TS architecture. An information decoder, an RF energy harvester and a switch to change the type of receiver antenna constitute the main components of the receiver in this architecture. The EH and ID inter-orbit receiving antenna or antennas are periodically changed. The time synchronization and the precise information/energy planning are can be considered as two essential parts in the receiver. The harvested energy by the $j_{th}$ receiver from the $i_{th}$ source in the EH mode can be defined as follows:
      \begin{equation}
          P_{i,j} = \eta P_{i} [h_{i,j}]^2
      \end{equation}
      where, $\eta$ is the efficiency coefficient of the EH process, $P_{i}$ is the transmission power of the $i_{th}$ source, and $h_{i,j}$ indicates to the gain of the channel that is implemented between the $i_{th}$ source and the $j_{th}$ receiver.

      The rate of the ID mode in the TS architecture is formulated as:
      \begin{equation}
          R_{i,j} = W log_2(1+\frac{P_{i} [h_{i,j}]^2}{N+I_j})
      \end{equation}
      where, $W$ is the transmission bandwidth, $N$ is the noise power, and $I_j$ is the interference signal power in receiver $j$.
  \item Power Splitting (PS): The PS receiver utilizes a unique approach to split the received signal into dual power streams, each with distinct power levels at a designated PS ratio prior to undergoing signal processing. Subsequently, these power streams are directed towards an information decoder and energy harvester, enabling simultaneous ID and EH application. Remarkably, this implementation integrates seamlessly with the existing PS architecture of conventional communication systems, demanding no further alterations except for the receiver circuit. Additionally, optimization of the PS ratio is feasible for each receiving antenna, offering the flexibility to strike a harmonious balance between information rate and harvested energy, in accordance with specific system requirements.

      Enhancing the overall performance involves optimizing signal mixing and PS ratios. We consider $\theta$ as the PS coefficient for receivers. $P_{i,j}$ is the energy that the $j_{th}$ receiver harvests from the $i_{th}$ transmitter and it can be formulated as:
      \begin{equation}
          P_{i,j} = \eta P_{i} [h_{i,j}]^2 \theta _j
      \end{equation}

      If we denote the noise power of the signal processing by $N_{sp}$, the ID rate of the communication link between the $i_{th}$ source and the $j_{th}$ receiver can be expressed as:
      \begin{equation}
          R_{i,j} = W log_2(1+\frac{(1- \eta )P_{i} [h_{i,j}]^2}{N_{sp}+N+I_j})
      \end{equation}

      Theoretically, it has been shown that the best compromise between the harvested energy and the information rate can be achieved in the PS architecture.
  \item Antenna Switching (AS): An antenna switch between the ID mode and the EH mode that s partially low complex enables the SWIPT.  The set of antennas in this architecture are divided into two subsets, one set of antennas is for harvesting energy and the other set is for transmitting information. The AS architecture is relatively simpler rather than the TS and the PS modes. Dual antenna can also be implemented in this architecture. In addition, the AS architecture can also be used to optimize a separate receiver architecture.
\end{itemize}

In Table 1, a comparison has been made between different states of EH [25].

\begin{table*}
\centering
\caption{EH Techniques}
\begin{tabular}{|c|p{120pt}|p{120pt}|}
  \hline
  Receiver Architecture & Advantages & Disadvantages \\
  \hline
  TS & Easy hardware implementation, practical for single antenna device & Coordination problem, need for proper schedule, occurrence of delay \\
  \hline
  PS & Applicable for single antenna device, ID and EH instant, less subject to delay & More complexity than TS, need to optimize PS coefficient \\
  \hline
  SR & Smaller hardware frame, single antenna transmitter & More complex architecture, subject to interference in the low power area \\
  \hline
  AS & Simultaneous implementation of EH and ID & Requirement of multiple antennas, subject to optimization error \\
  \hline
\end{tabular}
\end{table*}

Compared with AS and TS, PS is the best way to realize information transmission and power transmission at the same time. The RF signal collected by the antenna is divided based on the PS structure by a specific PS ratio $\% \theta$ where $\% \theta$ of the signal flows to the decoder circuit and $\% (1-\theta)$ simultaneously flows to the battery circuit.

In this paper, EH is used in a way that the extracted energy is used to charge the battery of the drone. In the proposed scenario, the UAV is a relay that harvests energy to be able to move to the second stop point. In general, two relay protocols can be implemented in UAV-aided wireless networks [26]:
\begin{itemize}
  \item Decode and Forward Relaying
  \item Amplify and Forward Relaying
\end{itemize}

In such scenarios, which are somehow the use of MIMO, several users are sending signals to the drone, and the drone continues to communicate with several users. In a MIMO system, the user terminals act as battery-limited devices, necessitating periodic recharging to prolong network longevity. Energy Harvesting (EH) is accomplished by individual receivers through dedicated power transmission from network transmitters. Prior research focused on integrating SWIPT into MIMO wireless networks by assuming distinct user groups: one for data reception and the other for energy replenishment. In fact, the main challenge is recharging the battery in a network that uses EH to harvest energy and then participate in communication [24].

To recharge the battery in a UAV that plays the role of a relay, we must pay attention to the architectural standards of the receiver, because the received signal is both a power signal and contains information. We assume that the relay or UAV in this research does not have a built-in power source, but it has a rechargeable energy storage device or battery. Therefore, the Bell relay harvests energy from the RF signals broadcast by the Uplink users, is powered, and operates in a save-then-cooperate (STC) mode [25].

The relay receives signals from any source and uses zero-forcing detecting (ZFD). Then the relay performs the network coding operation on the received signals and re-sends the resulting superposition of the signals to the destination. The destination has two antennas: one to broadcast the relatively high frequency EH signal (compared to its partner's antenna dedicated to receiving signals from other nodes) [26].

In the following, we will calculate the energy consumed in the drone and after that we will check how to harvest the energy and charge the battery of the drone.

The power that the UAV consumes is calculated as follows [11]:
\begin{equation*}
    P(v)=P_{0}(1+\frac{3v^{2}}{U_{tip}^{2}}\\)+P_{i}\sqrt{\sqrt{1+\frac{v^{4}}{4v_{0}^{2}}\\}\\-\frac{v^{2}}{2v_{0}^{2}}\\}\\+\frac{1}{2}\\d_{0}\rho sAv^{3}
\end{equation*}
Where $U_{tip}$, $v_{0}=\sqrt{\frac{W}{2\rho A}}$, $d_{0}$, $\rho$, $s$, and $A$ denote the tip speed of the rotor blade, the mean rotor induced velocity in hovering, the fuselage drag ratio, the air density, the rotor solidity, and the rotor disc area, respectively. $P_{0}=\frac{\delta}{8}\rho s A \Omega^{3} R^{3}$ and $P_{i}=(1+k)\frac{W^{\frac{3}{2}}}{\sqrt{2\rho A}}$ represent the blade profile power and the induced power when $v=0$. Where, $W$ is the UAV weight, $\Omega$ is the Blade angular velocity, $R$ denotes Rotor radius, and $k$ is the incremental correction factor to induced power.

The energy that the UAV consumes for NET1 and NET2 and also consumes to move from NET1 to NET2 is given as follows:
\begin{equation}
    E^{NET1}=P(0)\sum_{k\epsilon\rho_{c}^{N_{1}}}T_{k}^{UAV}\\
\end{equation}
\begin{equation}
    E^{NET2}=P(0) T_{BS}
\end{equation}
\begin{equation}
    E^{moving}=P(v_{moving})T^{moving}
\end{equation}
Where, $v_{moving}$ denotes the velocity of the UAV and $T^{moving}$ is the time that the UAV spends to move from NET1 to NET2 and will be calculated as:
\begin{equation}
    T^{moving}=\frac{d_{NET1}^{NET2}}{v_{moving}}\\
\end{equation}
Where, $d_{NET1}^{NET2}$ shows the distance between NET1 and NET2.

Applying the EH technique can be the main contribution of scenario B. As we said earlier, the PS receiver architecture is used in scenario B to provide the essential energy for the UAV. The UAV receives signals from $N_{CELLULAR}$ users from the first area in one time cycle, so the amount of energy that the UAV receives from the $k_{th}$ Uplink user in one cycle is calculated as follows:
\begin{equation}
    E_{UAV,k}^{EH}=\eta _{PS}^{EH} P_{R,k}^{UAV}
\end{equation}

Also, the amount of power spent on ID is given as follows:
\begin{equation}
    E_{UAV,k}^{ID}=(1- \eta_{PS}^{EH}) P_{R,k}^{UAV}
\end{equation}

The SNR of the communication of the $K_{th}$ cellular user with UAV in the scenario B can be defined as:
\begin{equation}
    \gamma_{k}^{UAV,S2} =\frac{(1- \eta_{PS}^{EH}) P_{R,k}^{UAV}}{n_{NET1}^{UAV}}\\
\end{equation}

As a result, the communication rate of the $K_{th}$ cellular user of the first area in scenario B in one cycle with the UAV is calculated as follows:
\begin{equation}
    \zeta_{k}^{UAV,S2}=Wlog_{2}(1+ \gamma_{k}^{UAV,S2})
\end{equation}

The total energy consumed by the UAV in one cycle from the first area to the second area can be defined as follows:
\begin{equation}
    E^{total}=E^{NET1}+E^{NET2}+E^{moving}
\end{equation}

This energy is the threshold energy required to maintain a constant transmission power in the UAV, $(E_{EH}^{th})$. Excess energy $(E_{EH}^{th})$ is stored in a Lithium ion battery, where the battery efficiency is denoted as $\eta^{Bat}$. This extra energy stored in the battery provides the basis for the next cycle. Finally, the energy level of the UAV battery is calculated as follows:
\begin{equation}
    E_{UAV}^{C+}=E_{UAV}^{C} + \eta^{Bat}((\sum_{k=1}^{N_{CELLULAR}} E_{UAV,k}^{EH})-E_{EH}^{th})
\end{equation}
where, $E_{UAV}^{C}$ is the energy that was available in the UAV's battery from the previous cycle, or it is the initial energy of the battery.

Finally, total time that the UAV spends will be given as:
\begin{equation}
    T^{total}=T^{moving}+T_{BS}+\sum_{k=1}^{N_{CELLULAR}}T_{k}^{UAV}\\
\end{equation}
In the next section, an optimization problem is formulated that optimizes the flight time of the UAV in scenario B.

\section{Optimization Problem}
The amount of charge in the drone's battery and the number of cellular users who are in line to communicate with the drone are the main factors affecting the completion of the drone's mission. Completing the UAV mission means that the UAV successfully receives the signals of cellular users from the first area and moves to the second area and sends data to the station located in the second area. According to the theoretical results of the previous sections and that the drone also receives energy from the first area, the duration of the drone's communication with each user is directly related to the amount of data received and the amount of energy received, but inversely related to the duration of the flight and inversely related to the energy It has used drones. As a result, the best network performance is achieved if a compromise is reached between the UAV flight time and the EH and battery characteristics of the UAV.

According to what was said before, the flight duration of the drone is directly defined by the amount of energy in the drone's battery and the amount of energy received and consumed. The duration of the flight consists of three basic parts, the duration of the drone's communication with the users of the first area, the duration of the movement of the drone from the first area to the second area, and the duration of the communication with the station of the second area. During the time that the drone is in contact with the users of the first zone, it must serve all the users, as a result, serving all the users of the first zone is one of the limitations of the problem of optimizing the flight time of the drone. Next, the UAV must have enough energy to reach the second area and send data to the base station of the second area.
\begin{equation}
    opt.prob.
    \begin{cases}
        \begin{bmatrix}
            \eta^{Bat},\ \
            \eta_{PS}
        \end{bmatrix}^{opt} =  argmin\ \ T^{total}
        \\ \\ c_1 : \ \  E_{UAV}^{C+}\ge E_{EH}^{th}
        \\ \\ c_2 : \ \  0 \le \eta^{Bat} \le 1
        \\ \\ c_3 : \ \  0 \le \eta_{PS} \le 1
    \end{cases}
\end{equation}

After some calculations, the optimization problem is simplified to 27.

\begin{table*}
\begin{equation}
    opt.prob.
    \begin{cases}
        \begin{bmatrix}
            \eta^{Bat},\ \
            \eta_{PS}
        \end{bmatrix}^{opt} =  argmin\ \ \frac{1}{\gamma_{BS}^{UAV}}\\ + \sum_{k=1}^{N_{CELLULAR}}\frac{1}{\gamma_{k}^{UAV,S2}}
        \\ \\ c_1 : \ \  E_{UAV}^{C} + P_{T}^{NET_{1}} \eta _{PS}^{EH} \eta^{Bat}(\sum_{k=1}^{N_{CELLULAR}} R_{k,UAV}^{-\alpha _{u}}) - (1+\eta^{Bat})E_{EH}^{th}\ge 0
        \\ \\ c_2 : \ \  0 \le -\eta^{Bat} + 1
        \\ \\ c_3 : \ \  0 \le -\eta_{PS} + 1
    \end{cases}
\end{equation}
\end{table*}

Proof of 27: See Appendix 1.

This minimization problem, equation 27, can be transformed to a maximization problem as is shown in 28.

\begin{table*}
\begin{equation}
    opt.prob.
    \begin{cases}
        \begin{bmatrix}
            \eta^{Bat},\ \
            \eta_{PS}
        \end{bmatrix}^{opt} =  argmax\ \ \gamma_{BS}^{UAV} + \sum_{k=1}^{N_{CELLULAR}}\gamma_{k}^{UAV,S2}
        \\ \\ c_1 : \ \  E_{UAV}^{C} + P_{T}^{NET_{1}} \eta _{PS}^{EH} \eta^{Bat}(\sum_{k=1}^{N_{CELLULAR}} R_{k,UAV}^{-\alpha _{u}}) - (1+\eta^{Bat})E_{EH}^{th}\ge 0
        \\ \\ c_2 : \ \  0 \le -\eta^{Bat} + 1
        \\ \\ c_3 : \ \  0 \le -\eta_{PS} + 1
    \end{cases}
\end{equation}
\end{table*}

7, 8, 21 and 22 can be substituted at 28 to reach equation 29.

\begin{table*}
\begin{equation}
    opt.prob.
    \begin{cases}
        \begin{bmatrix}
            \eta^{Bat},\ \
            \eta_{PS}
        \end{bmatrix}^{opt} =  argmax\ \ \frac{P_{T}^{UAV}H^{-\alpha _{u}}}{n_{NET2}^{UAV}}\\ + \sum_{k=1}^{N_{CELLULAR}}\frac{(1- \eta_{PS}^{EH}) P_{T}^{NET_{1}}R_{k,UAV}^{-\alpha _{u}}}{n_{NET1}^{UAV}}\\
        \\ \\ c_1 : \ \  E_{UAV}^{C} + P_{T}^{NET_{1}} \eta _{PS}^{EH} \eta^{Bat}(\sum_{k=1}^{N_{CELLULAR}} R_{k,UAV}^{-\alpha _{u}}) - (1+\eta^{Bat})E_{EH}^{th}\ge 0
        \\ \\ c_2 : \ \  0 \le -\eta^{Bat} + 1
        \\ \\ c_3 : \ \  0 \le -\eta_{PS} + 1
    \end{cases}
\end{equation}
\end{table*}

It is assumed that the transfer power of the UAV is a constant that depends on the energy of the UAV's battery, so equation 29 can be simplified to equation 30.

\begin{table*}
\begin{equation}
    opt.prob.
    \begin{cases}
        \begin{bmatrix}
            \eta^{Bat},\ \
            \eta_{PS}
        \end{bmatrix}^{opt} =  argmax\ \ \sum_{k=1}^{N_{CELLULAR}}\frac{(1- \eta_{PS}^{EH}) R_{k,UAV}^{-\alpha _{u}}}{n_{NET1}^{UAV}}\\
        \\ \\ c_1 : \ \  \eta _{PS}^{EH} \eta^{Bat} G1 - E_{EH}^{th} \eta^{Bat} + G2 \ge 0
        \\ \\ c_2 : \ \  0 \le -\eta^{Bat} + 1
        \\ \\ c_3 : \ \  0 \le -\eta_{PS} + 1
    \end{cases}
\end{equation}
\end{table*}

In equation 30, $G2= E_{UAV}^{C} - E_{EH}^{th}$ and $G1 = P_{T}^{NET_{1}} (\sum_{k=1}^{N_{CELLULAR}} R_{k,UAV}^{-\alpha _{u}})$ are constant values.

It is clear from 30 that the object function and constraints of the optimization problem are linear convex. Hence, the optimization problem can be solved by primary or dual problem.

The Lagrange function is obtained as shows in equation 31 to satisfy the Karush–Kuhn–Tucker (KKT) conditions.

\begin{table*}
\begin{equation}
    \begin{cases}
    L(\eta^{Bat}, \eta_{PS}, \lambda_{1}, \lambda_{2}, \lambda_{3}) = -\sum_{k=1}^{N_{CELLULAR}}\frac{(1- \eta_{PS}^{EH}) R_{k,UAV}^{-\alpha _{u}}}{n_{NET1}^{UAV}}\\ -\lambda_{1}(\eta _{PS}^{EH} \eta^{Bat} G1 - E_{EH}^{th} \eta^{Bat} + G2) - \lambda_{2}(-\eta^{Bat} + 1)\\ - \lambda_{3}(-\eta_{PS} + 1)
    \end{cases}
\end{equation}
\end{table*}

We use $(\eta^{Bat})^*$ and $(\eta_{PS})^*$ as optimal values of $\eta^{Bat}$ and $\eta_{PS}$, so the following point is an stationary point for the Lagrange function:
$((\eta^{Bat})^*, (\eta_{PS})^*, (\lambda_{1})^*, (\lambda_{2})^*, (\lambda_{3})^*)$

Hence, the partial derivative of 31 with respect to $\eta^{Bat}$ and $\eta_{PS}$ is taken as follows:
\begin{equation}
    \begin{cases}
    \frac{\partial L(\eta^{Bat}, \eta_{PS}, \lambda_{1}, \lambda_{2}, \lambda_{3})}{\partial \eta^{Bat}}\\  = -\lambda_{1}(\eta _{PS}^{EH} G1 - E_{EH}^{th}) + \lambda_{2}
    \end{cases}
\end{equation}
\begin{equation}
    \begin{cases}
    \frac{\partial L(\eta^{Bat}, \eta_{PS}, \lambda_{1}, \lambda_{2}, \lambda_{3})}{\partial \eta_{PS}}\\  = -\sum_{k=1}^{N_{CELLULAR}}\frac{-R_{k,UAV}^{-\alpha _{u}}}{n_{NET1}^{UAV}}\\ -\lambda_{1} \eta^{Bat} G1  + \lambda_{3}
    \end{cases}
\end{equation}

32 and 33 have to be equated to 0, so:
\begin{equation}
    \eta _{PS}^{EH} = \frac{\lambda_{2} + \lambda_{1}E_{EH}^{th}}{\lambda_{1} G1}
\end{equation}
\begin{equation}
  \eta^{Bat} = \frac{\lambda_{3} -\sum_{k=1}^{N_{CELLULAR}}\frac{-R_{k,UAV}^{-\alpha _{u}}}{n_{NET1}^{UAV}}\\}{-\lambda_{1} G1}
\end{equation}

Following that, the partial derivative of 32 with respect to $\lambda_{1}$, $\lambda_{2}$ and $\lambda_{3}$ is calculated as follows:
\begin{equation}
    \begin{cases}
    \frac{\partial L(\eta^{Bat}, \eta_{PS}, \lambda_{1}, \lambda_{2}, \lambda_{3})}{\partial \lambda_{1}}\\  = \eta _{PS}^{EH} \eta^{Bat} G1 - E_{EH}^{th} \eta^{Bat} + G2
    \end{cases}
\end{equation}
\begin{equation}
    \begin{cases}
    \frac{\partial L(\eta^{Bat}, \eta_{PS}, \lambda_{1}, \lambda_{2}, \lambda_{3})}{\partial \lambda_{2}}\\  = -\eta^{Bat} + 1
    \end{cases}
\end{equation}
\begin{equation}
    \begin{cases}
    \frac{\partial L(\eta^{Bat}, \eta_{PS}, \lambda_{1}, \lambda_{2}, \lambda_{3})}{\partial \lambda_{3}}\\  = -\eta_{PS} + 1
    \end{cases}
\end{equation}

Considering the conditions of KKT and the fact that 37, 38 and 39 have to be equated to 0, following equation is obtained:
\begin{equation}
    \eta _{PS}^{EH} \eta^{Bat} G1 + G2 = E_{EH}^{th} \eta^{Bat}
\end{equation}

After some mathematical manipulations, the optimal value of the EH parameters can be defined as follows:
\begin{equation}
    (\eta^{Bat})^* = \frac{G2}{E_{EH}^{th}}
\end{equation}
\begin{equation}
    (\eta_{PS})^* = \frac{G1 E_{EH}^{th}}{G2}
\end{equation}

The theoretical results show that the scenario proposed in this article using EH techniques is an efficient way to empower the UAV to complete its mission. Next, the simulation results show that in the second scenario, where EH is used for drones, more users can be served. In addition, by determining the number of specific users for the second scenario, the optimal duration of the UAV flight is also evaluated in the simulation results.

\section{Simulation Results}
In this section, some features of investigated DRN and the UAV, namely rate, harvested energy from uplink transmissions, consumed energy by the UAV to serve whole users are studied. Studying the effect of the DRN’s parameters such as the density of users of NET1, UAV altitude, threshold of transferred data to the UAV, the transfer power of uplink users and transfer power of the UAV on the performance of the DRN is the main purpose of this section. Essential parameters for simulation are listed in table1.
\begin{table}
\centering
\caption{Simulation Parameters}
\label{table}
\setlength{\tabcolsep}{25pt}
    \begin{tabular}{|p{25pt}|p{75pt}|}
        \hline
        Parameter&
        Value \\
        \hline
        $U_{tip}$&
        {120} m/s \\
        \hline
        $v$&
        70 m/s \\
        \hline
        $d_{0}$&
        {0.6} \\
        \hline
        $\rho$&
        {1.225} kg/m$^{3}$ \\
        \hline
        $\alpha _{u}$&
        {2} \\
        \hline
        $s$&
        {0.05} \\
        \hline
        $A$&
        0.503 m$^{2}$ \\
        \hline
        $R$&
        0.4 meter \\
        \hline
        $k$&
        0.1 \\
        \hline
        $\omega$&
        13 radian/seconds \\
        \hline
        $\delta$&
        0.012 \\
        \hline
        $W$&
        20 Newton \\
        \hline
        $n_{NET2}^{UAV}$&
        0 dB \\
        \hline
        $H^{UAV}$&
        800 m \\
        \hline
        $n_{NET1}^{UAV}$&
        0 dB \\
        \hline
    \end{tabular}
\end{table}

Figure (2) indicates the complete agreement between the simulation results and the theoretical results for the total spent time by the UAV to serve all cellular users on NET1 and move to the second area to send signals to the BS of this area. It is obvious that the total spent by the UAV decreases with the increase of transfer power of uplink users because by increasing the transfer power, the rate of uplink users increases. By increasing the communication rate, the UAV will receive the specified amount of data in a shorter period of time, and the amount of energy harvested will also increase.
\begin{figure}[!t]
    \centering
    \includegraphics[width=6.5in,trim={0 6cm 0 6cm},clip]{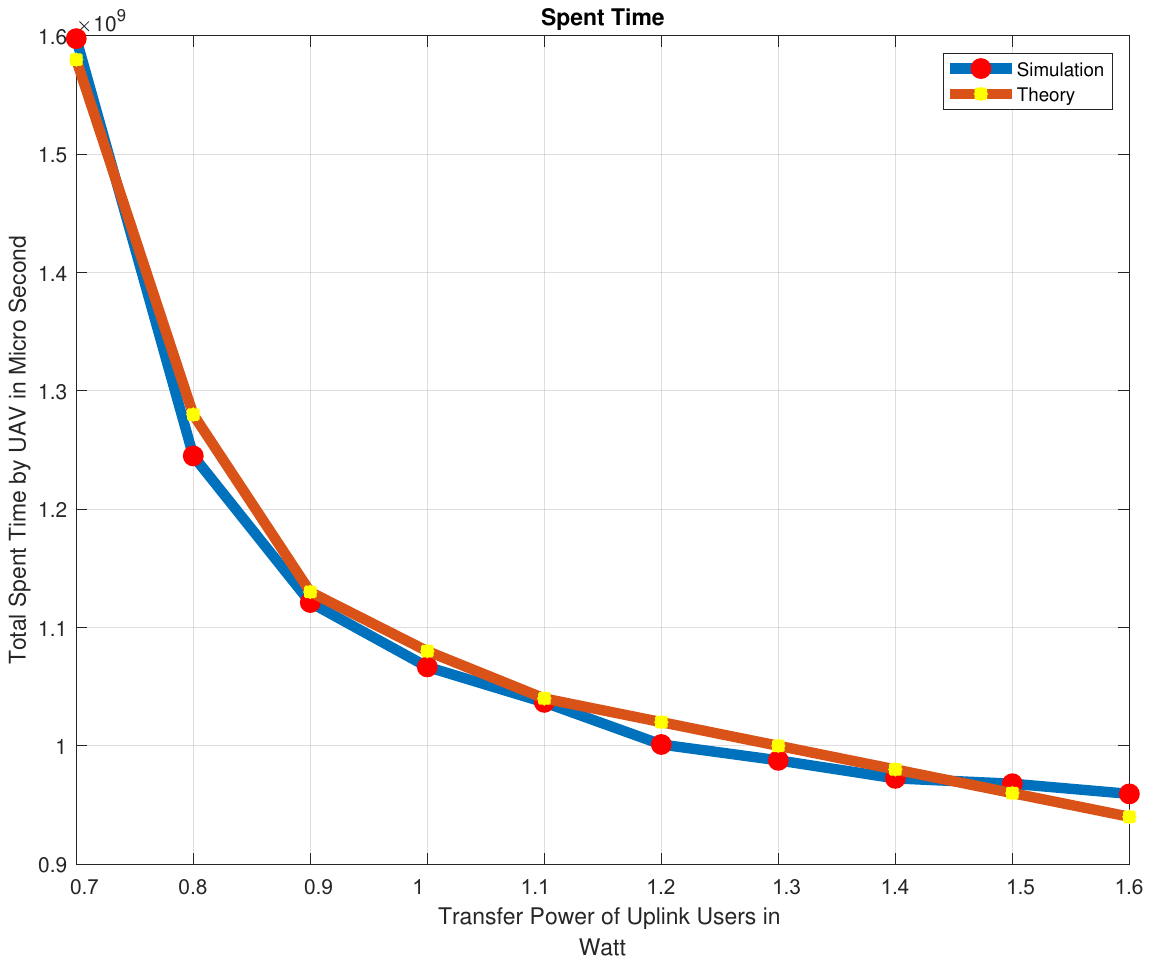}
    \caption{spent time by the UAV vs. Transfer Power of Uplink Users }
    \label{fig2}
\end{figure}

Figure (3) shows the effect of the density of uplink cellular users of NET1 on the rate of uplink communication in the first scenario. We use $P_{T}^{NET_{1}} = [1\ \ 3\ \ 5] Watt$. It can be seen from figure (3) that the total rate of uplink transmissions grows over density of uplink users, also it increases by growing the transfer power of users. As the uplink transmissions are orthogonal and uplink users do not have interference for each other, by increasing the total number of users, the total rate of uplink communications grows. Moreover, there is a direct relation between the transfer power of users and the rate of the uplink communication.
\begin{figure}[!t]
    \centering
    \includegraphics[width=6.5in,trim={0 6cm 0 6cm},clip]{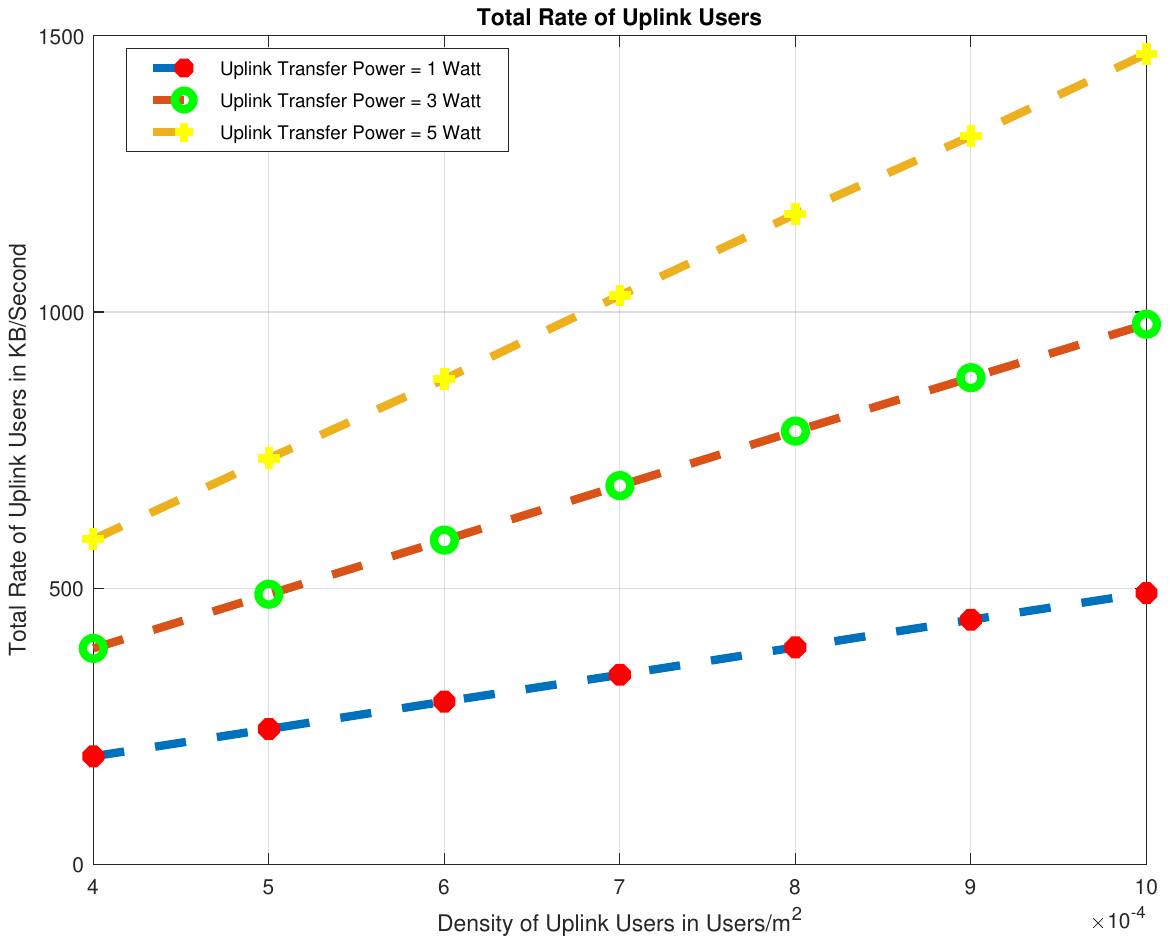}
    \caption{Total Rate of Uplink Transmission vs. Density of Uplink Cellular Users}
    \label{fig3}
\end{figure}

The effect of changing the sending power of users in area 1 on the number of users that the UAV can serve in the second scenario is examined in Figure (4). The more users can send, the better the communication rate and the drone receives more energy, so the user can serve more users. Considering this figure and comparing with the first scenario where the number of users is limited due to the energy of the drone, it is clear that in the second scenario we can serve more users with the benefit of EH techniques. Despite the fact that increasing the sending power leads to more cost, but in the second scenario of this article, we have been able to make the most of the consumed power by using EH techniques.
\begin{figure}[!t]
    \centering
    \includegraphics[width=6.5in,trim={0 6cm 0 6cm},clip]{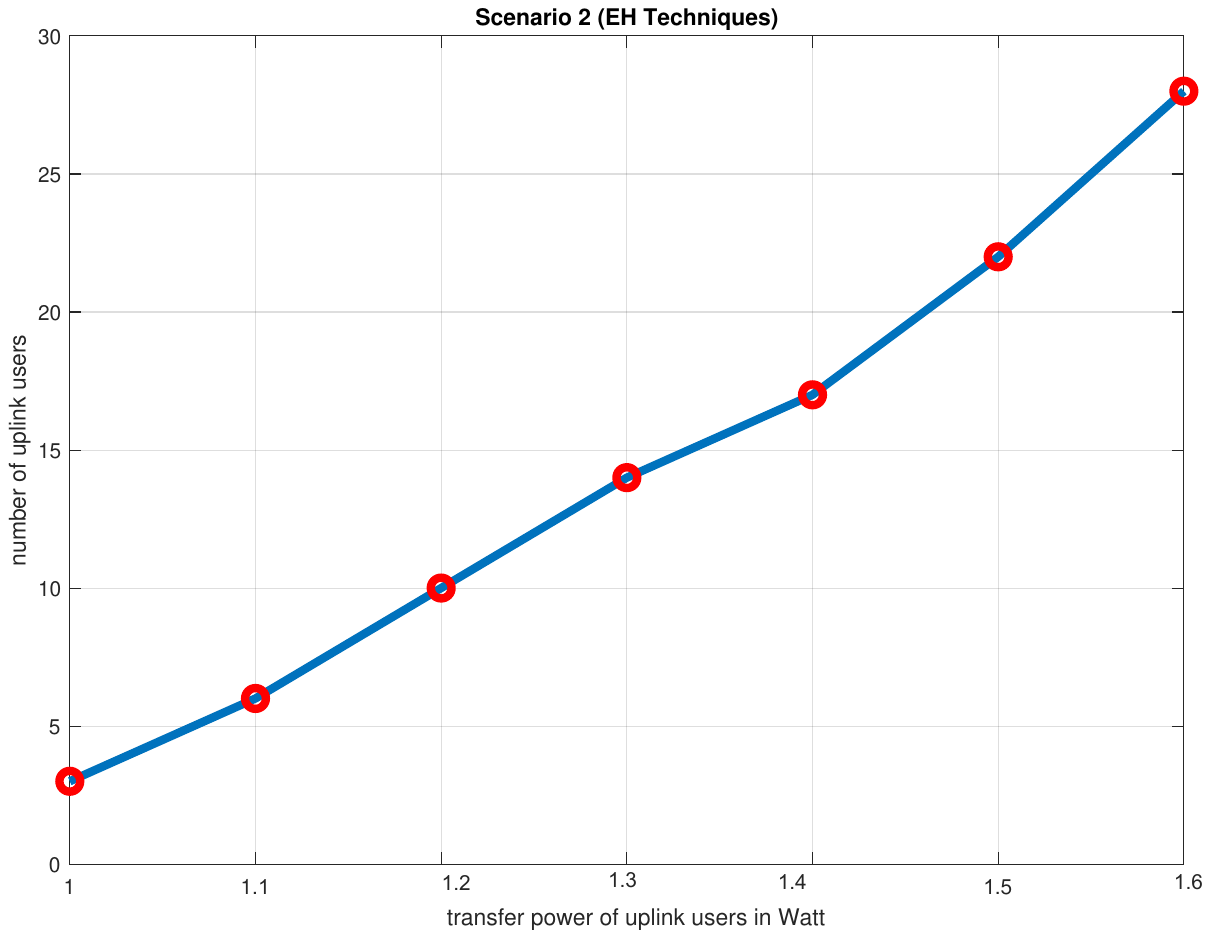}
    \caption{Total number of uplink users vs. Transfer power of uplink users in Watt}
    \label{fig4}
\end{figure}

Figures (5) and (6) illustrate the impact of the threshold of data that the UAV can collect from uplink users of the NET1 and send to the users of NET2 on the total consumed energy by the UAV and total spent time by the UAV in the second scenario respectively. It is assumed that $P_{T}^{NET_{1}} = {5} Watt$, $P_{T}^{UAV} = {3} Watt$ and $H^{UAV} = [500 \ 700 \ 900]$. It is clear from figure 5 that the total consumed power by the UAV goes up with increase of the threshold data of the UAV. In addition, it is obvious from figure 6 that there is an increase in the spent time by the UAV by growing the threshold data. In both figures 5 and 6, an upward trend can be seen for the consumed power and spent time by incresing the altitude of the UAV. As expected, when the UAV hovers in higher altitude, it will need more time to collect and send data.
\begin{figure}[!t]
    \centering
    \includegraphics[width=6.5in,trim={0 6cm 0 6cm},clip]{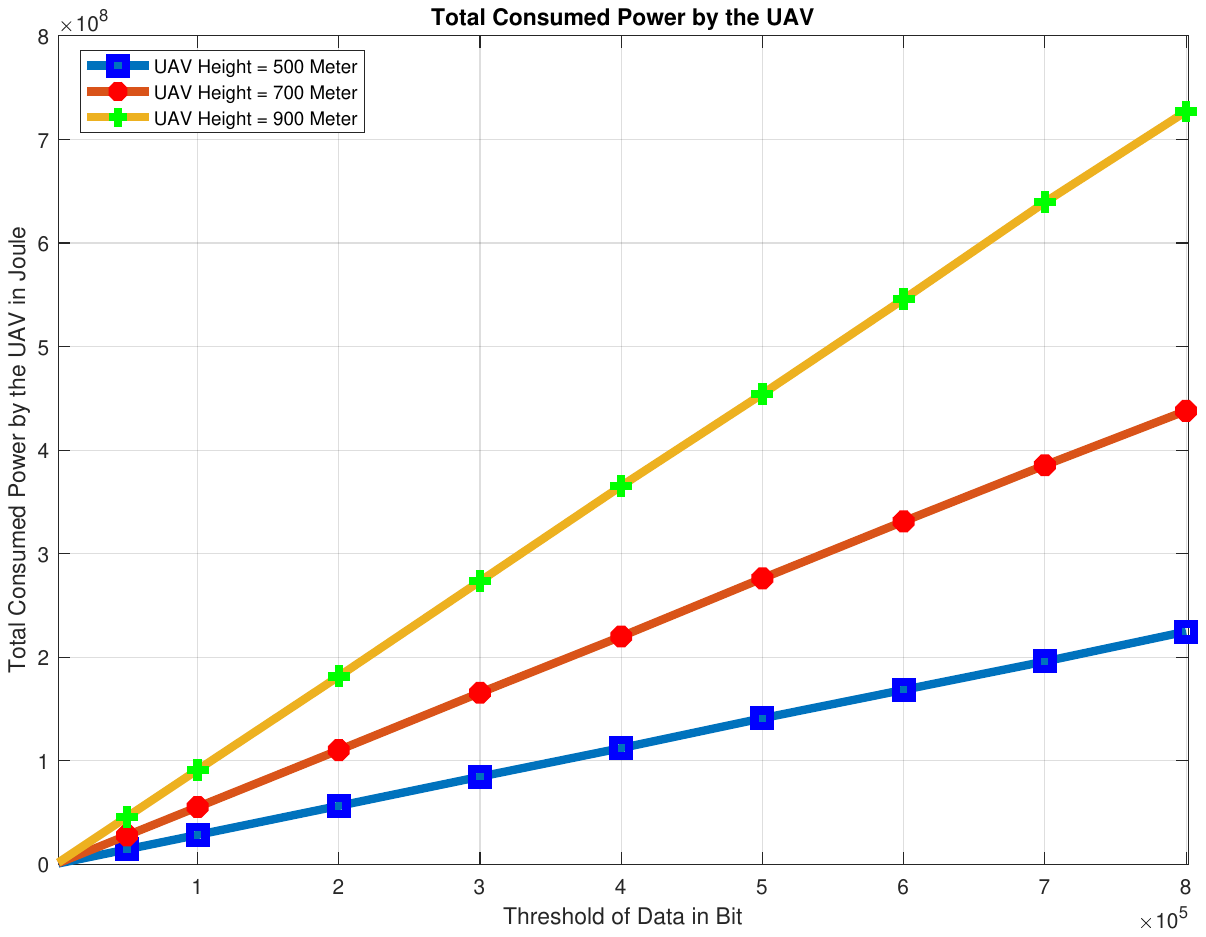}
    \caption{Total Consumed Energy by the UAV vs. Threshold of Data}
    \label{fig5}
\end{figure}
\begin{figure}[!t]
    \centering
    \includegraphics[width=6.5in,trim={0 6cm 0 6cm},clip]{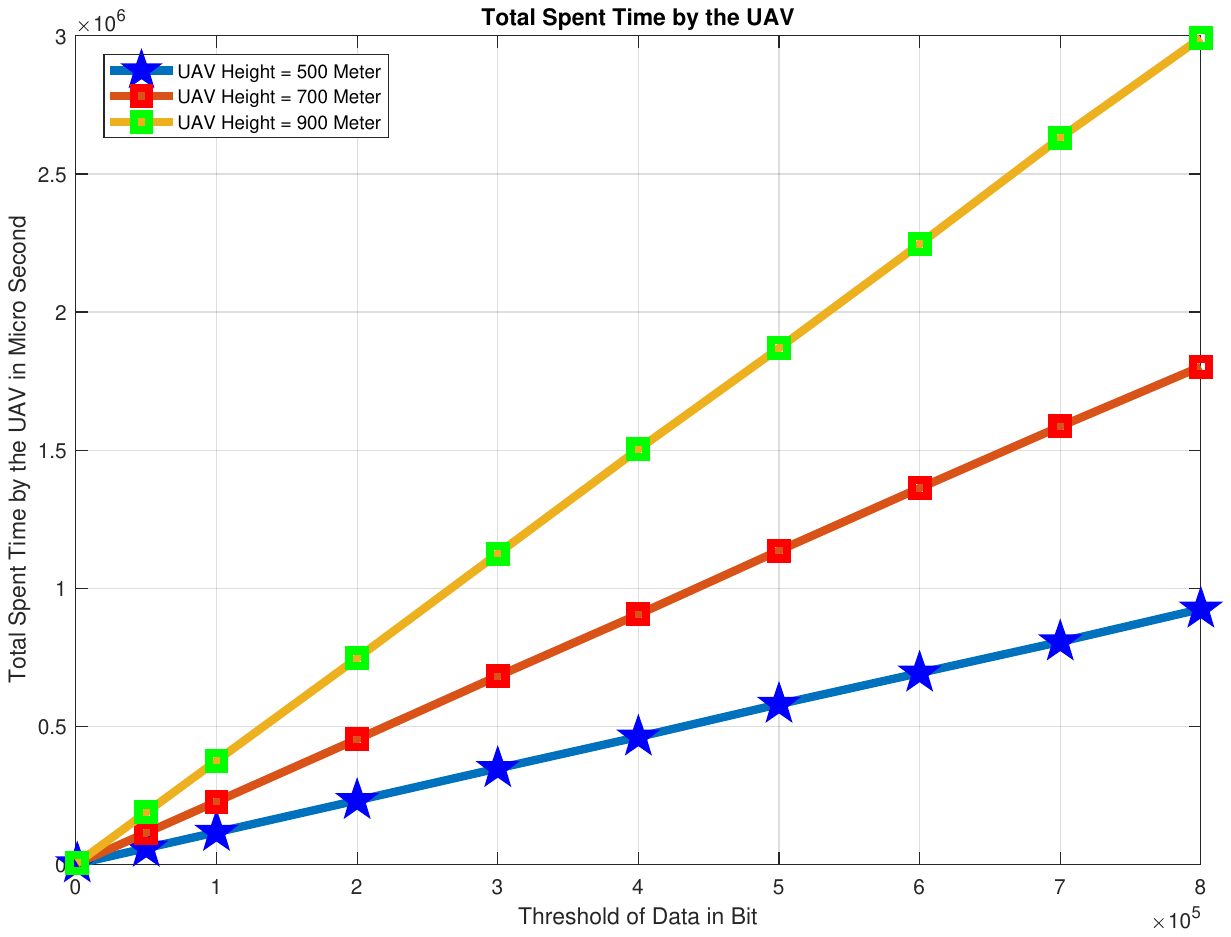}
    \caption{Total Spent Time by the UAV vs. Threshold of Data}
    \label{fig6}
\end{figure}

Figure (7) studies the effect of the density of uplink users of NET1 on the total spent time by the UAV in the second scenario. We use $H^{UAV} = {700}$ and $D_{th}^{UAV} = 4 * 10^5 Bits$. It can be seen from figure 7 that the total spent time by the UAV decreases at first with the increase of the density of uplink users, and then grows with more increase of the density. At first, as the density is relatively small, the amount of the harvested energy by the drone is insignificant compared to the total consumed energy. After a certain level of the density, the total harvested energy begins to grow because of the severe increase of the harvested energy.
\begin{figure}[!t]
    \centering
    \includegraphics[width=6.5in,trim={0 6cm 0 6cm},clip]{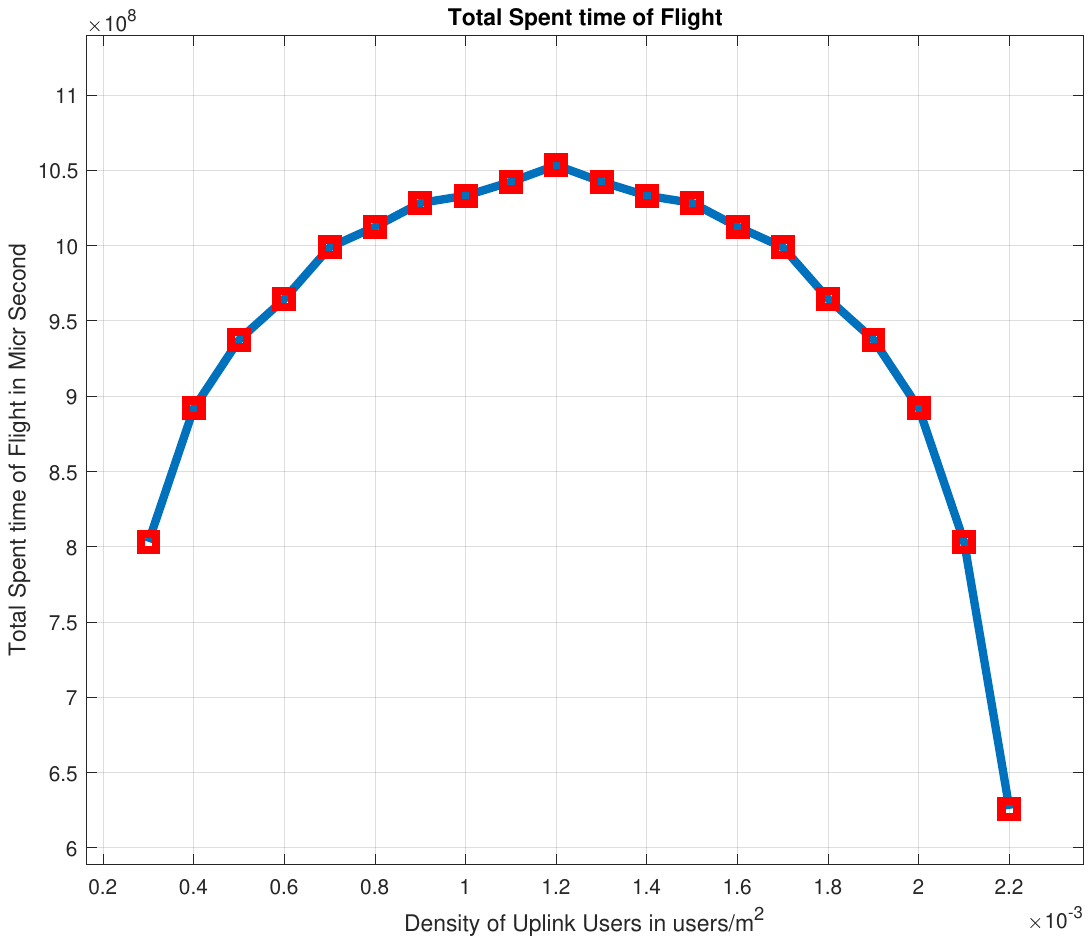}
    \caption{Flight time vs. Density of Uplink Users}
    \label{fig7}
\end{figure}

The relation between the transfer power of cellular users of NET1 and the altitude of the UAV is shown in figure (8). It is assumed that $D_{th}^{UAV} = 4 * 10^5 Bits$. It is clear from figure 8 that when the transfer power of users declines, the UAV will hover in lower altitude to serve all users. Hence, the useful altitude of the UAV decreases with reducing the transfer power. Moreover, when the users's density increases, the UAV is forced to hover in a lower altitude.
\begin{figure}[!t]
    \centering
    \includegraphics[width=6.5in,trim={0 6cm 0 6cm},clip]{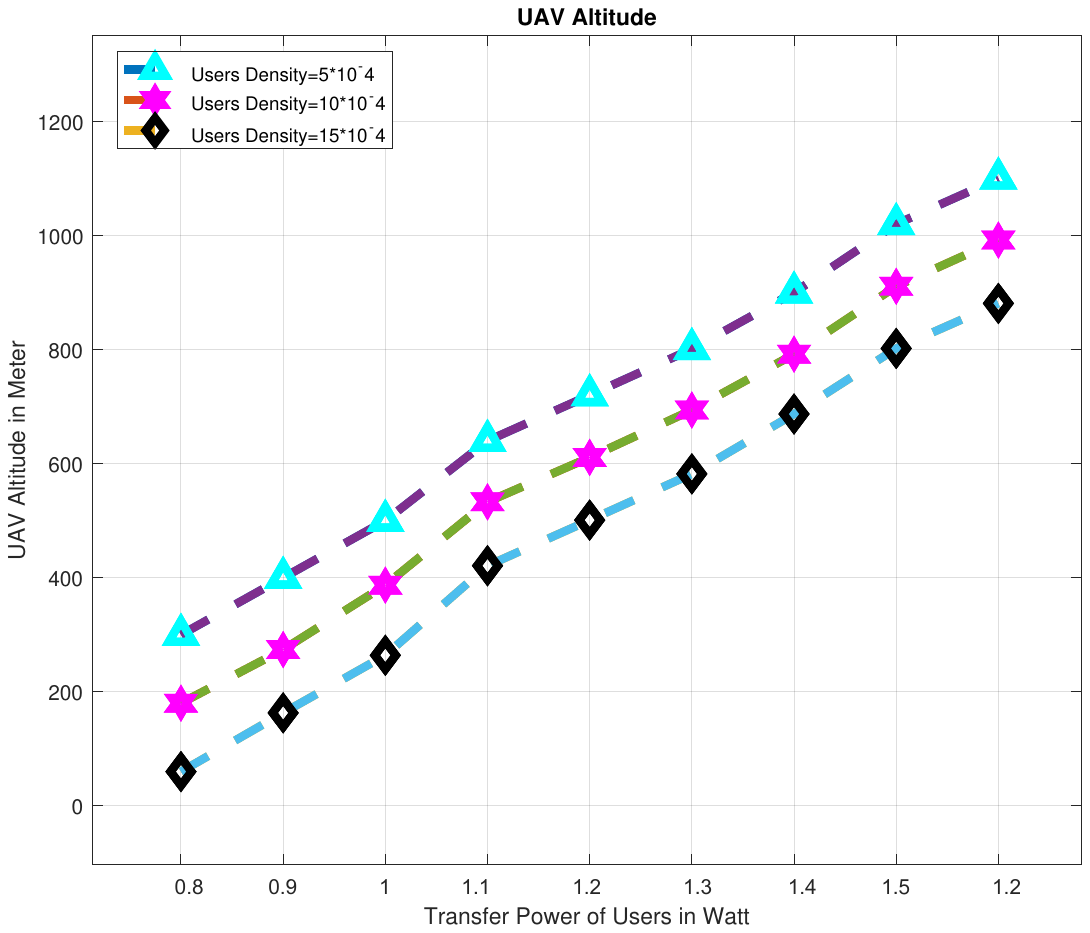}
    \caption{Altitude of the UAV vs. Transfer Power of The Users of NET2}
    \label{fig8}
\end{figure}

Figure (9)
\begin{figure}[!t]
    \centering
    \includegraphics[width=6.5in,trim={0 6cm 0 6cm},clip]{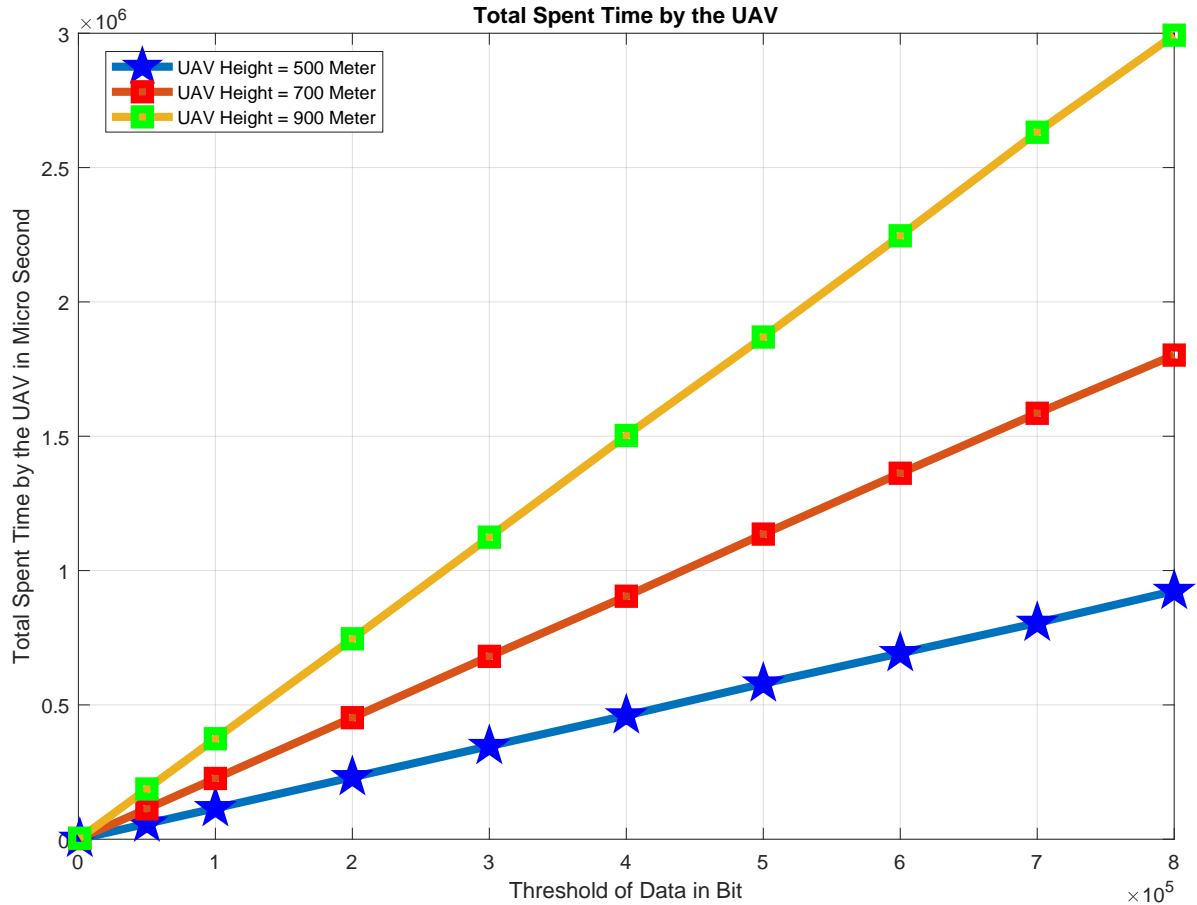}
    \caption{Total Spent Time by the UAV vs. Threshold of Data}
    \label{fig6}
\end{figure}

\section{Conclusion}
This work has studied the system performance of an UAV-enabled DRN assited by SWIPT technology. By receiving sig-nals from users involved in a DRN and transmitting informa-tion to another area, the UAV plays the role of a relay between two areas. Two different scenarios are compared in this paper and the main difference of which is that in the second scenario, SWIP technology is used to design a communication protocol between users involved in the DRN and the UAV. Based on the comparison of the two scenarios, it is clearly shown that by using SWIPT techniques, the network efficiency increases significantly. An important achievement that this efficiency improvement shows is that the number of users from the disaster area receiving service increases dramatically, which is clearly shown in the simulation and theory results. In future work, a network of IOT users in an area that will form a multiple-input multiple-output (MIMO) system with several UAVs assisted by beamforming and SWIPT technologies can be studied.

The main perspective for future work in the field discussed in this paper can be seen as a compromise between network communication delay, network energy efficiency and data security. By using machine learning techniques, a suitable solution can be found to achieve this compromise. In fact, future work can be directed more towards training users and drones to save energy and ensure data security.

\section{Appendix 1}
By Substituting 24 and 25 at 26, we reach to equation 42.

\begin{table*}
\begin{equation}
    opt.prob.
    \begin{cases}
        \begin{bmatrix}
            \eta^{Bat},\ \
            \eta_{PS}
        \end{bmatrix}^{opt} =  argmin\ \ T^{moving}+T_{BS}+\sum_{k=1}^{N_{CELLULAR}}T_{k}^{UAV}
        \\ \\ c_1 : \ \  E_{UAV}^{C} + \eta^{Bat}(\sum_{k=1}^{N_{CELLULAR}} E_{UAV,k}^{EH}) - (1+\eta^{Bat})E_{EH}^{th}\ge 0
        \\ \\ c_2 : \ \  0 \le \eta^{Bat} \le 1
        \\ \\ c_3 : \ \  0 \le \eta_{PS} \le 1
    \end{cases}
\end{equation}
\end{table*}

Equations 6, 10, 19 and 20 can be substituted at 27 to reach to equation 43.

\begin{table*}
\begin{equation}
    opt.prob.
    \begin{cases}
        \begin{bmatrix}
            \eta^{Bat},\ \
            \eta_{PS}
        \end{bmatrix}^{opt} =  argmin\ \ \frac{1}{\zeta_{BS}^{UAV}}\\ + \sum_{k=1}^{N_{CELLULAR}}\frac{1}{\zeta_{k}^{UAV}}
        \\ \\ c_1 : \ \  E_{UAV}^{C} + \eta _{PS}^{EH} \eta^{Bat}(\sum_{k=1}^{N_{CELLULAR}} P_{R,k}^{UAV}) - (1+\eta^{Bat})E_{EH}^{th}\ge 0
        \\ \\ c_2 : \ \  0 \le -\eta^{Bat} + 1
        \\ \\ c_3 : \ \  0 \le -\eta_{PS} + 1
    \end{cases}
\end{equation}
\end{table*}

By substituting 1, 9, 22  at 28, we reach to equation 44.

\begin{table*}
\begin{equation}
    opt.prob.
    \begin{cases}
        \begin{bmatrix}
            \eta^{Bat},\ \
            \eta_{PS}
        \end{bmatrix}^{opt} =  argmin\ \ \frac{1}{Wlog_{2}(1+ \gamma_{BS}^{UAV})}\\ + \sum_{k=1}^{N_{CELLULAR}}\frac{1}{Wlog_{2}(1+ \gamma_{k}^{UAV,S2})}
        \\ \\ c_1 : \ \  E_{UAV}^{C} + P_{T}^{NET_{1}} \eta _{PS}^{EH} \eta^{Bat}(\sum_{k=1}^{N_{CELLULAR}} R_{k,UAV}^{-\alpha _{u}}) - (1+\eta^{Bat})E_{EH}^{th}\ge 0
        \\ \\ c_2 : \ \  0 \le -\eta^{Bat} + 1
        \\ \\ c_3 : \ \  0 \le -\eta_{PS} + 1
    \end{cases}
\end{equation}
\end{table*}

Note that it can be assumed that $(P_{LOS,k}^{NET_{1}}+\eta P_{NLOS,k}^{NET_{1}})=1$.

Finally,29 can be simplified to equation 45.

\begin{table*}
\begin{equation}
    opt.prob.
    \begin{cases}
        \begin{bmatrix}
            \eta^{Bat},\ \
            \eta_{PS}
        \end{bmatrix}^{opt} =  argmin\ \ \frac{1}{\gamma_{BS}^{UAV}}\\ + \sum_{k=1}^{N_{CELLULAR}}\frac{1}{\gamma_{k}^{UAV,S2}}
        \\ \\ c_1 : \ \  E_{UAV}^{C} + P_{T}^{NET_{1}} \eta _{PS}^{EH} \eta^{Bat}(\sum_{k=1}^{N_{CELLULAR}} R_{k,UAV}^{-\alpha _{u}}) - (1+\eta^{Bat})E_{EH}^{th}\ge 0
        \\ \\ c_2 : \ \  0 \le -\eta^{Bat} + 1
        \\ \\ c_3 : \ \  0 \le -\eta_{PS} + 1
    \end{cases}
\end{equation}
\end{table*}

\end{document}